\documentclass[12pt]{article}            
\usepackage{graphicx}
\usepackage{multicol}
\begin{document}
\centerline{\large{\bf Serendipity in Astronomy}}
\bigskip
\centerline{\large{A.C. Fabian}}
\centerline{Institute of Astronomy, University of Cambridge, UK}
\vspace{1cm}

\noindent Astronomy is an observationally-led subject where chance discoveries
play an important role. A whole range of such discoveries is
continually made, from the trivial to the highly significant. What is
generally needed is for luck to strike someone who is prepared, in the
sense that they appreciate that something novel has been
seen. ``Chance favours the prepared mind'' in the words of Pasteur (1854). 

This is one definition of serendipitous discovery, first identified as
such by Horace Walpole in a letter in 1754 to Horace Mann on
discussing a Persian tale of three Princes of Serendip. We shall hear
Several more interpretations\footnote{see e.g. The Travels and
Adventures of Serendipity, Merton \& Barber 2006} are outlined in
these chapters, but I shall stick with the concept of a chance or
unplanned discovery. In contrast with school laboratory science where
the aim is to plan and carry out an experiment in controlled
conditions, in general astronomers cannot do this and must rely on
finding something or a situation which suits. Often, the possibilities
afforded by a phenomenon are only appreciated later, after the
surprise of the discovery has worn off.

A nice example is the discovery in 1979 of the volcanoes of Io by
Voyager 1.  This phenomenon was spotted by a navigational engineer,
Linda Morabito, rather than one of the project scientists. Io turns
out to be the most volcanically-active body in the Solar System.

\begin{figure}[h]
\hbox{\centerline{
\includegraphics[width=0.5\textwidth,angle=0]{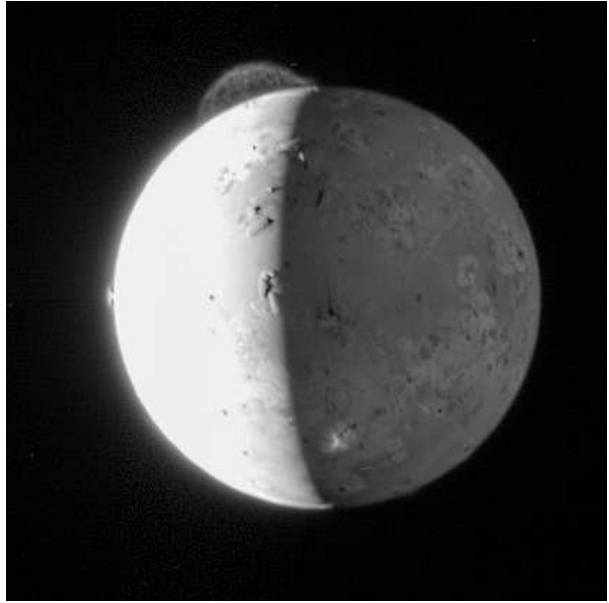}}}
\caption{Several active volcanoes can be seen in this NASA image of
  Io. This moon is tidally squeezed and heated by Jupiter, which leads
  to the eruptions. Since Io has little atmosphere the ejecta reach
  high altitudes before raining back. }
\end{figure}

`Chance favours the prepared mind' implies both an element of luck and
a prior understanding of what is normal. Making successful discoveries
in astronomy is not comparable to buying a lottery ticket and then
sitting back but requires a deep familiarity with the Sky, the
Universe, cosmic phenomena and/or physics. It does require both sides;
you don't make discoveries without making observations and you don't
identify them as such without knowing when something is new. As some
examples from chemistry clearly demonstrate, some clumsiness, or at
least a deviation from what would otherwise be the path of best
practice, may also be needed (mercury was discovered to be a catalyst
for synthetic indigo when a mercury thermometer was broken in the
reaction vessel; Roberts 1989).  It may not prove easy to build
algorithms or robots to make serendipitous discoveries!

I am sure that chance plays a strong part in the way in which we
learned the world as children. Playing is just that and we all start
out with curiosity outweighing prejudice. Only later, alas, does
prejudice come to dominate and become the enemy of discovery. It must
have played an important role in the development of astronomy and
science. It is possible of course to continue to make serendipitous
(if not original) discoveries oneself, such as why the full moon
always rises at sunset. 

An illustration of how I define serendipity is shown in Figure 1. The
axes of this 3-dimensional figure are luck, preparedness and aim. Pure
serendipity lies just on the luck--preparedness plane, whereas the
perfectly-planned, Kantian (Glashow 2002), experiment is just on the
preparedness--aim plane. In reality many serendipitous
discoveries have some aim (Archimedes was surely thinking about the
problem of estimating the volume of an irregular object when he
stepped in his bath), and I shall allow that. There just has to be a
large unexpected element.

\begin{figure}
\hbox{\centerline{
\includegraphics[width=0.5\textwidth,angle=270]{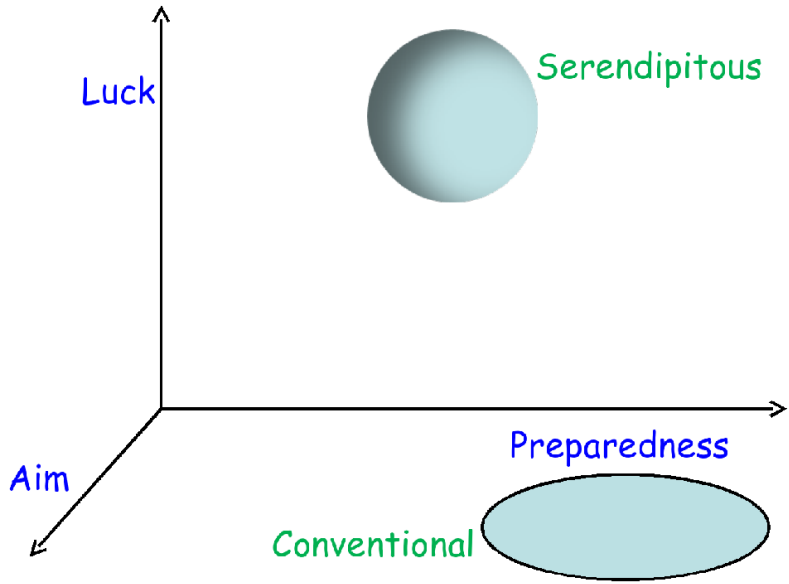}}}
\caption{Serendipitous discoveries combine luck (or chance),
  preparedness and aim. Conventional school science is usually
  concentrated on the preparedness, aim plane and totally new things
  are found on the preparedness, luck plane. There is usually, however,
  some aim to the observations, which is why I have tried through
  shading to make the region have volume. }
\end{figure}

One way to categorize discoveries is through Discovery Space (Figure 2,
Harwit 1982). New things are generally found when new parts of
discovery space are explored. We need to go more than about 3 times
deeper in that space to find new things. Discovery Space is
multi-dimensional and not just in space and time but in spectral band
(e.g. radio vs visible), spectral resolution (discerning different
frequencies or colours), time resolution, polarization etc. A familiar
analogy to a multi-dimensional space is provided by our senses;
hearing is distinct from smell or sight and if we make them more acute
we discover new things. Imagine having the sensitivity to smell of a
dog, or the sensitivity to electrical activity of a shark!

\begin{figure}
\hbox{\centerline{
\includegraphics[width=0.5\textwidth,angle=270]{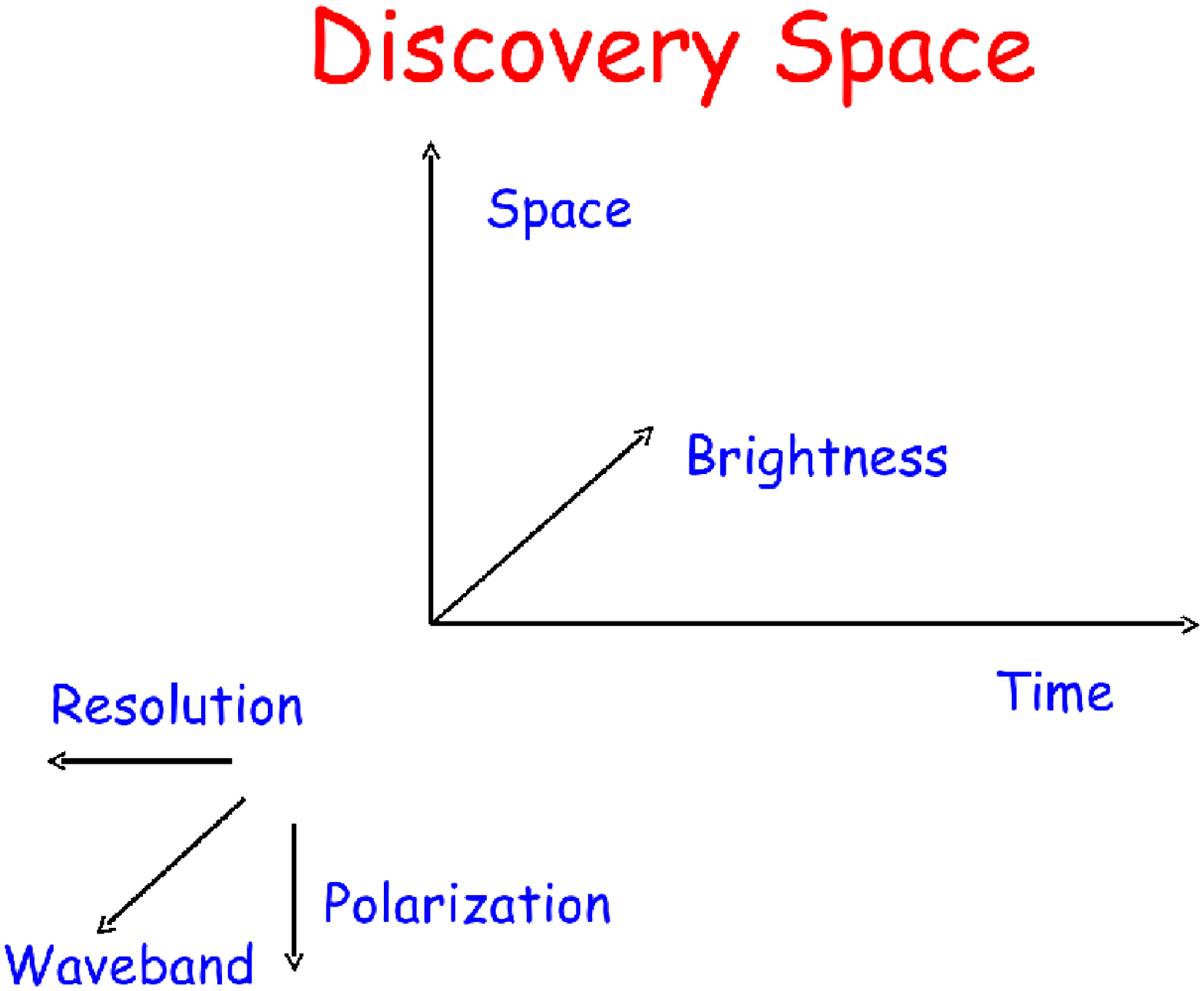}}}
\caption{A schematic representation of some parts of discovery
  space. The axes could represent the volume of space surveyed, the
  brightness level down to which the observations are sensitive and
  the epoch of the observations. Alternatively they could be spatial
  resolution or contrast level, or ability to discern rapid variations.}
\end{figure}

Serendipity often involves an element of surprise, which implies an
emotional quality to scientific work. Like a good joke which leads in
one direction before jumping to another, so a serendipitous discovery
-- a Eureka, or even a `what the..'  moment can jump our train of
thought to new directions. The way in which a serendipitous discovery
turns our thoughts, either individually or collectively, into new
directions is one of the major benefits of serendipity.

Note that discovery doesn't always make you famous. Stigler's law of
eponyms states that ``no scientific discovery is named after its
original discover'' and indeed Stigler did not discover that law
either. It is also worth stressing that there is lots of noise or
spurious signals in most observational searches so it pays to become
very familiar with the instruments and how they work.

Astronomy is powered by serendipitous observations and serendipity is
an accepted mode of progress. There are hundreds of research papers
with `serendipitous' in the title and I found 17 in 2007. One theme
which I shall develop is that although serendipity is part of
astronomy it is not directly factored in to the decision or funding
processes to the extent that it might successfully be. Science's
outside appearance of being rational and methodological is somewhat at
odds with serendipity. There is a tension associated with it and a
`fishing expedition' is a derogatory term used by time assignment and
funding panels. This may be a legacy of our innate feeling that
serendipity is somehow obvious, random or childish.

I have spent much of my career in X-ray astronomy, which had a
serendipitous beginning. Indeed, X-rays were discovered
serendipitously by Ro\"entgen in 1895. We are all familiar with X-rays
for their property of being easily absorbed by the bones of our
body. The astronomical X-rays mostly observed by astronomers are even
more easily absorbed and can only travel a few inches in
air. Consequently, X-ray astronomy must be carried out above the
Earth's atmosphere from high-altitude balloon, rocket or satellite.
Solar X-rays were discovered in the 1950s (see Friedman 1990) but it
was estimated that it would be impossible to detect the X-rays from
other stars using the equipment of the time. Riccardo Giacconi
proposed to NASA using a sounding rocket to look for solar X-rays
reflected from the Moon. In a short rocket flight on 18 June 1962 he
and his team discovered diffuse X-ray emission from around the Sky and
a peak of emission in a direction which did not point at the
Moon. They had found Sco X-1 and the X-ray Background (XRB). We now
know that the first is due to matter from a normal star falling onto
an orbiting neutron star; the XRB is due to matter falling into
distant supermassive black holes.

Unexpected discoveries are routinely made in astronomy, even sometimes
by amateurs. The NASA website Astronomy Picture of the Day, which is
one of the first things I log into each day, has most of them and can
certainly be appreciated by everyone.  A recent example, Comet
Holmes, first discovered in 1872, flared up by a factor of 500,000
last November and was first spotted by J.A. Henriquez Santana in
Tenerife as a naked-eye object in the constellation of Perseus. Why it
flared up is still a mystery, although the list of possible reasons is
growing. It is unlikely to have been due to a collision with something
else since it probably did the same abrupt brightening when it was
discovered. It must have an inherently unstable core; presumably it
briefly had much stronger `volcanoes' than Io.  Such events in the Sky
must have been familiar to the Chinese and Korean court astronomers
who watched the Sky as a way to predict the future of both the Emperor
and the State. Some of the events they witnessed were supernovae,
which mark the collapse and subsequent explosion of stars. One such
led to the expanding supernova remnant known as the Crab Nebula (it
doesn't really look like a Crab, but then the constellation of Cancer
doesn't either). We now know that the solid remnant of the collapsed
star is a neutron star spinning at 30 times a second. Neutron stars
were predicted to exist in the 1930s soon after neutrons were
discovered in Cambridge. They have the mass of the Sun but a radius of
just 15~km, which is smaller than London. Long suspected to be the
power source of the nebula, the object was not identified as such
until after pulsars were serendipitously discovered by Jocelyn Bell,
Anthony Hewish and others in 1967.

\begin{figure}
\hbox{\centerline{
\includegraphics[width=0.5\textwidth,angle=0]{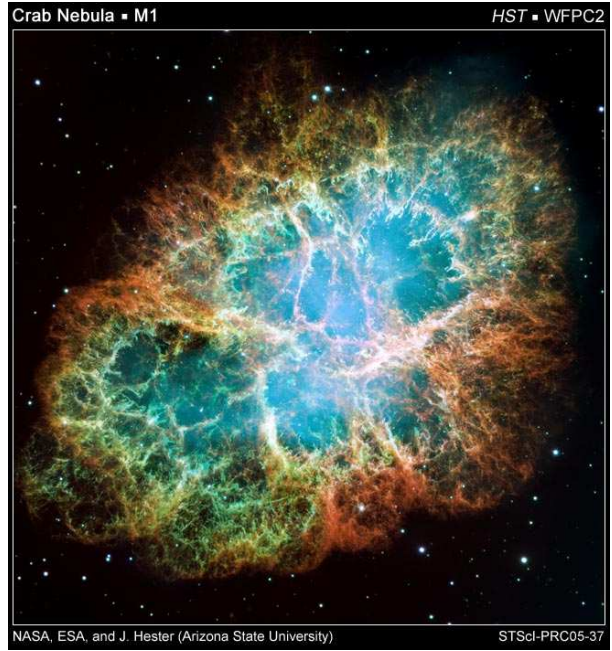}}}
\caption{The Crab nebula, which is about 6000 light years away. It is
  the remnant of a massive star  seen to explode in
  1054. High energy electrons from the pulsar near the centre create
  the blue glow.}
\end{figure}

They were looking for the flickering of so-called radio stars, due to
the solar wind, in order to determine how small they are (the effect
of turbulence in our atmosphere on the visible light from stars which
are pointlike makes them twinkle, whereas planets which are extended
do not). They built an array of aerials sensitive to variations of a
second or less. Jocelyn Bell discovered `scruff' every sidereal day on
the pen recorder charts within which she found 1.3~s pulses. Little
Green Men as the responsible agent were ruled out when several more
pulsars were found. They are rapidly spinning, highly magnetized
neutron stars of which the Crab pulsar is just one, which played a key
role in developing the theory of how they operate since the nebula
acts as a calorimeter for the power radiated.

Interestingly the pulsar was suspected to be the energy source before
(it was known as Baade's star). I've heard stories that the occasional
lay observer saw the object twinkling (their eyes may have been
particularly sensitive to variations). X-ray data of the Crab taken
just before the Cambridge radio discovery of pulsars was later found
to show pulses but, and this is important, was not analysed in a way
that revealed the pulses at the time (Fishman et al 1969). How many
important discoveries are lying in someone's archive or hard disk?

Sometimes, supernovae mark the formation of a black hole, which can
produce intense, brief, gamma-ray emitting jets seen as a gamma-ray
burst (GRB). Observations of GRB have been dominated by serendipitous
discoveries from the very start when seen by US military (Vela)
satellites in 1967 and announced in 1972. Some are observed right
across the Universe and are the brightest things in the Universe
during their brief life. Some GRB are different and are due to quakes
on neutron stars with superstrong magnetic fields, one thousand times
stronger than those of a typical radio pulsar and several thousand million
million times stronger than the Earth's magnetic field in a room.
Such objects are called magnetars and a recent outburst on 27 December
2004 from SGR\,1806 gave the strongest event ever observed at any
wavelength. For the first 2 tenths of a second the (mostly gamma-ray)
energy flux on Earth received from this object, which is about 30,000
light years away, exceeded that of a Full Moon. It was then
intrinsically one thousand times brighter than the 100 billion stars
in our Milky Way galaxy! It saturated all gamma-ray detectors and some
of the best measurements of its maximum brightness came from a Russian
satellite instrument which was in Earth's shadow and saw the emission
reflected from the Moon. The decay phase of the event consisted of 5
sec pulsations which were even detected as oscillations in Earth's
magnetic field (the gamma-rays ionized the upper atmosphere which in
turn affected the magnetic field).  If astrologers wanted to have a
scientific basis for their `predictions' they would take up gamma-ray
astronomy. If the event had been much closer then it could have
profoundly affected us all -- in a very negative manner.

\begin{figure}
\hbox{\centerline{
\includegraphics[width=0.5\textwidth,angle=0]{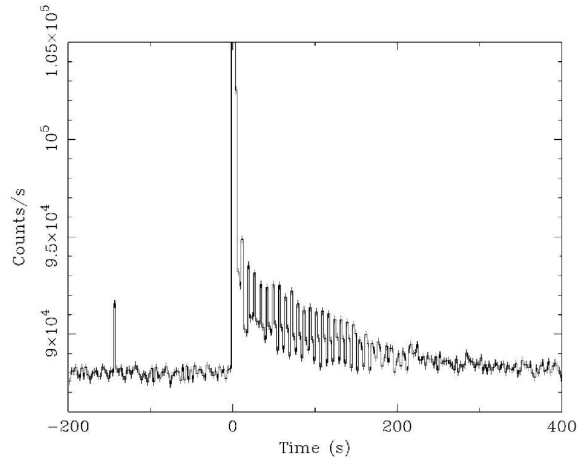}}}
\caption{Gamma-ray light curve of the soft gamma-ray repeater SGR1806
  from December 2004. The spike near t=0 is saturated. Note the
  oscillations in the decay phase. which caused the Earth's magnetic
  field to oscillate. }
\end{figure}

The time domain is the least explored one in astronomy and continues to
be rich in discoveries. It is likely to be opened up further over the
next decade with telescopes mapping the visible Sky every few days
such as PanStarrs, LSST, Lofar, Gaia, SKA etc. Such instruments will
produce vast amounts of data (many Terabytes) every day so
analysing the data will be a serious challenge. This is becoming an
issue with the enormous facilities now coming on line in many
subjects. The Large Hadron Collider is an example from physics. So
much data is produced that most has to be eliminated
immediately. How to optimize such enormous data gathering exercises for
serendipitous discoveries is unclear.

Sometimes the object or effect was predicted earlier, but considered
too faint or difficult to be seen. This was the case with both
neutron stars and black holes which are so tiny that it seemed
reasonable to assume that they would be unobservable. What was
required was for the emission to be stimulated (as in a laser) for the
pulsar or highly beamed by relativistic outflow for the GRB. Any
scientific paper written predicting their observational detection well
before they were discovered would rightly have been rejected!

The discovery of the black hole at the Galactic Centre is a further
example of this. While not exactly a serendipitous discovery, it was
made possible by the lucky, and still unexplained occurrence of
suitable markers (the He stars) in that place. Two teams, one led by
Genzel in Munich the other by Ghez in Los Angeles, have seen and
followed the motion of individual bright stars orbiting the central
black hole. One star has now been seen over the past 15 years to
complete one whole orbit. The precision with which this elliptical
orbit has been measured leaves us in no doubt that the central 4
million solar masses of our Galaxy is extremely compact and can only
be a black hole.

One effect of gravity is to bend light. This was first measured for
the Sun during a total solar eclipse by Cambridge astronomer Arthur
Eddington in 1919 and as it matched Einstein's predictions it is what
made Einstein famous. Further examples were lacking for almost 60 years
during which time few people considered the effect from an
observational point of view. Then in 1979 a double quasar\footnote{A
  quasar is an accreting massive black hole producing so much
  radiation that it outshines its host galaxy.}  was discovered by
Walsh, Carswell and Weymann, who were making routine measurements of
the properties of a large sample of quasars.  Two of the quasars were
separated by 6 arcsec and were found to have identical spectra. So
identical that at first they didn't think that the telescope operator
had actually moved the telescope. What they had discovered was light
being deflected above and below a massive galaxy along the line of
sight. When things are exactly lined up the background source appears
as a ring (an Einstein ring) around the lensing galaxy, but in a more
typical case such as the double quasar where things are slightly
mismatched then the image appears as two separate objects (a third one
occurs at the centre but is often absorbed by dust in the lensing
galaxy).  Both images of the double quasar vary with what we now know
is the same pattern yet shifted in time by 430 days, which is the
difference in light travel time along the two paths.  Gravitational
lensing has now become (and remains) something of an astronomical
industry. More dramatic examples were found in clusters of galaxies in
the 1980s. Again these were found serendipitously. They had certainly
been seen earlier (and can be seen in published images) but not
noticed as such.

\begin{figure}
\hbox{\centerline{
\includegraphics[width=0.5\textwidth,angle=0]{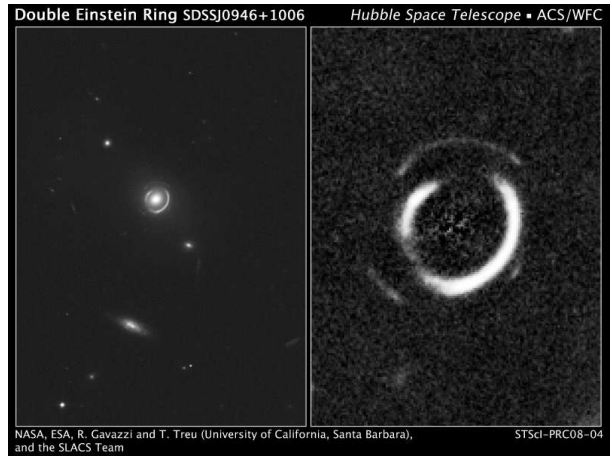}}}
\caption{A double Einstein ring caused by gravitational lensing. Three
  galaxies lie almost exactly in a line, with the nearest one bending
  the light from the more distant ones into two rings of light.}
\end{figure}
\begin{figure}
\hbox{\centerline{
\includegraphics[width=0.5\textwidth,angle=0]{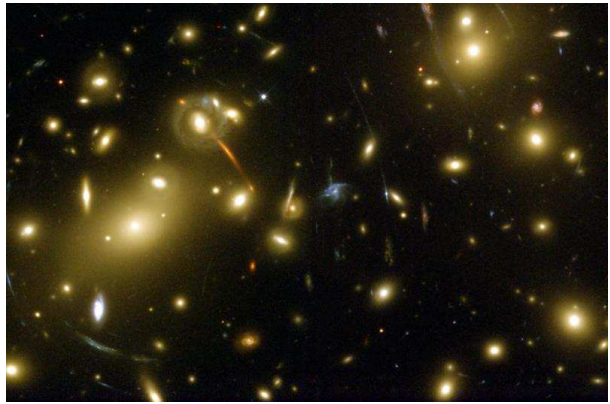}}}
\caption{The large fluffy objects are galaxies in a rich cluster, the
  total mass of which is gravitationally bending the light from
  distant galaxies into many arcs. In essence the whole cluster core,
  several hundred thousand light years across, acts as a giant
  telescope.}
\end{figure}

One of my own serendipitous discoveries is of ripples in the hot gas
at the centre of a cluster. To observe the gas requires X-rays and the
image was taken with the Chandra X-ray telescope.  The ripples
correspond to quasi-spherical ripples in the pressure of the gas. They
are strong sound waves created by the action of gas accreting into the
supermassive black hole at the centre of the central galaxy of the
cluster. There are several notable properties of these sound waves,
the first is that they carry lots of energy to large radii, in essence
they enable the central black hole to have a significant influence on
gas over intergalactic distances, the second is that they have a very
low frequency of one ripple per 10 million years!  It corresponds to a B
flat about 57 octaves below middle C.  The main response of the UK
media to our press release on this was just the B flat!

\begin{figure}
\hbox{\centerline{
\includegraphics[width=0.5\textwidth,angle=0]{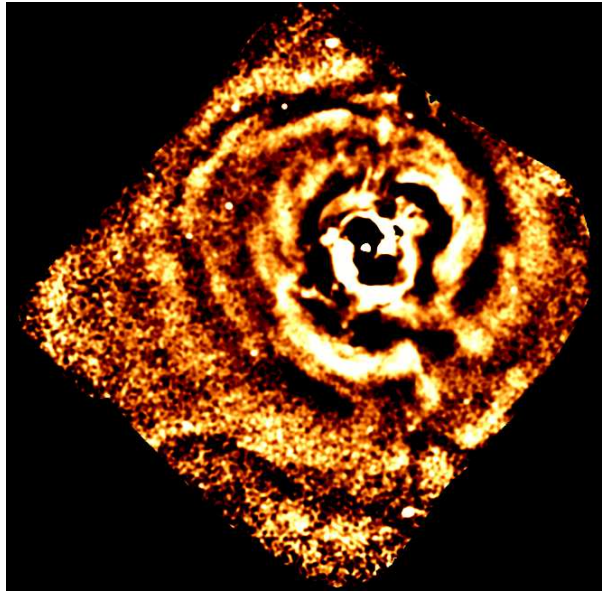}}}
\caption{Chandra X-ray image of the centre of the Perseus cluster of
  galaxies. The X-rays are produced by hot gas lying between the
  galaxies. Jets squirting from close to a massive black hole at the
  centre blow a sequence of bubbles in the gas which create sound
  waves seen as ripples in the gas.  }
\end{figure}

As my final example I return to our Galaxy to discuss the case of
extrasolar planets, which were long sought for but only found orbiting
normal stars in 1995 by Mayor and Queloz. (Curiously, 3 planetary-mass
objects were found orbiting a pulsar in 1991.) The problem lay in
everyone's expectations which supposed that other solar systems would
resemble our own. In our own Solar system the most massive planet,
Jupiter, orbits far out every 12 years creating a very tiny response
in our Sun which would be difficult to measure in other stars. In the
first extrasolar system found, it was a Jupiter-mass planet in a 4 day
orbit! Being so close means that its effect on the star is much larger
than if it were in a 12 year orbit, swinging that star about at 60
rather than 12.5 m per sec. Such `hot jupiters' give a much larger
signal than anyone had expected. Now hundreds have been found, not all
with such extreme orbits, but nevertheless few of them actually
resembling our own (many of much more highly eccentric orbits than
found in our Solar System). 55 Cancri is the nearest one with 5
planets.

This leads me to an area of research in astronomy which relies solely
on serendipity, namely the search for extraterrestrial
intelligence. What are the chances of detecting another intelligent
civilization? Drake's equation gives us some idea. The number of
civilizations, $N$, in the Galaxy with which we can communicate is
given by $N=Rf_pnf_lf_if_cL,$ where $R$ is the rate with which stars
are being formed (about 1 per year), $f_{\rm p}$ is the fractions of
stars which have planets, $n$ is the number of those planets which are
habitable and $f_l, f_i$ and $f_c$ are the fractions of those planets
on which life, intelligence and communicable civilizations occur. $L$
is the lifetime in years of a communicable civilization.  The
detection of extrasolar planets has pinned down one of the very
uncertain numbers, $f_p,$ (to say 10\% or more). Even then the
optimist will find that $N\sim0.001-0.01 L$.  A key unknown is the
length of time a technologically aware civilization lasts, $L$. Ours
is only just over 100 years old and we need them to last millions of
years for $N$ to be in the 1000s or more. If they do last that long
then we are most likely to find a civilization in the middle of its
run, not just near the start as we are. Therefore we will be dealing
with civilizations which are hundreds of thousands to millions of
years more technologically advanced than our own.

\begin{figure}
\hbox{\centerline{
\includegraphics[width=0.8\textwidth,angle=0]{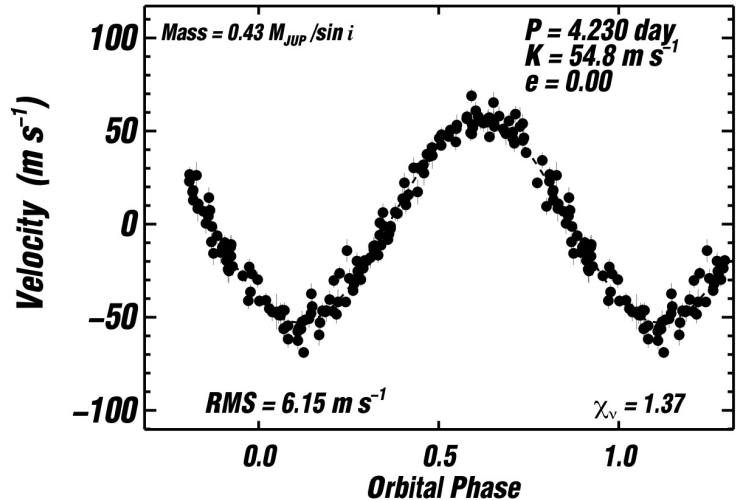}}}
\caption{Velocity of the star 51 Peg along our line of sight (courtesy
 of   G. Marcy). It
  oscillates by about 55 m/s due to a Jupiter-mass planet orbiting it
  every 4 days.}
\end{figure}

If so, then what will they look like and how will they communicate? Would
our present methods seems as absurd for interstellar communication as
smoke signals would do across continents? Should we look for highly
directed radio signals, which is where the effort has gone so far, or
maybe for flashes of X-rays or gamma-rays, as I suggested 30 years
ago (Fabian 1977)? Won't any intelligent civilization be looking for
serendipitous flashes in the night?

It is clear that astronomy is rich in serendipitous discovery and
phenomena. Harwit predicted that such discoveries would soon slow down
as most things were discovered. His argument was based on the
assumption that the phenomena have equal weight but given that they
are not I foresee such discoveries continuing on well into the
future. Several of my examples have been drawn from the past few
months and years and even then I had a wide range of discoveries to
choose from. In many areas, astronomy is still in an exploratory
discovery phase.  Cosmology too is equally rich, as is discussed
by Simon Singh. In the near future we can expect surprises from
neutrino and gravitational wave astronomy. We may yet find that dark
matter, which comprises 21\% of the Universe, comes in 42 different
varieties.

Does it not then make sense to tailor our research funding, both for
the hardware -- the telescopes -- and for the modes of working, to
take this into account? This leads to a current dilemma. Facilities
(and people) are increasingly expensive. Funding agencies using public
funds want value for money so are most likely to fund projects and
telescopes and teams where a successful outcome is predicted. This
tends to mean looking into areas close to where we know, rather than
stepping out into the unknown. In the case of serendipitous discovery
there is little that can be predicted with certainty, we can only
argue on the basis of past success. It means stepping out boldly in
discovery space.

This situation contrasts with laboratory physics where experiments are
mostly controlled and conditions predictable. This is not to say that
serendipitous discovery is then absent, as is discussed by
Richard Friend, but when a new facility costs millions to billions of pounds
then it is generally the clearest defined case with predictable
outcomes which wins the money. The situation with respect to astronomy
has become polarized by an emphasis on fundamental physics in some
quarters. In that view, how a galaxy works is of less interest than
the more fundamental issue of whether there are or are not other
dimensions to space, for example. Astronomy is often tied up in the
messy complexity of everything, whereas the `fundamentalists' want to
study the (assumed) much cleaner basis of it all.

\begin{figure}
\hbox{\centerline{
\includegraphics[width=0.5\textwidth,angle=270]{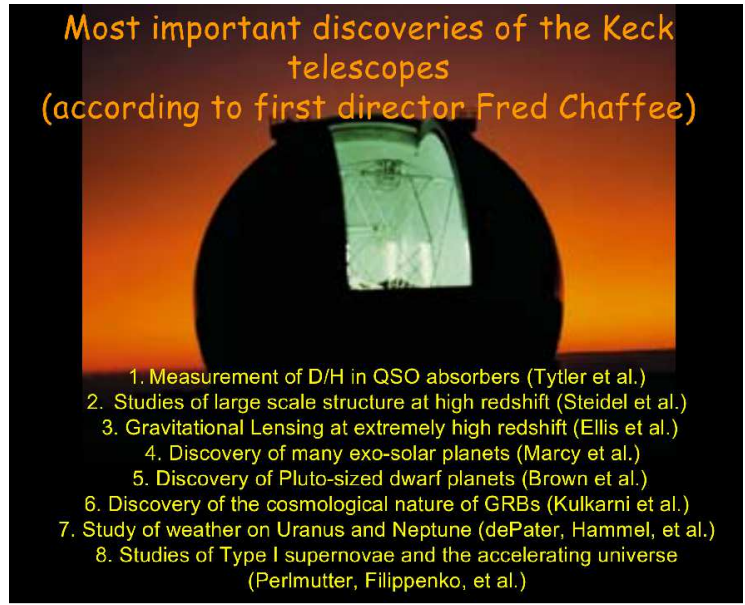}}}
\caption{Some Keck discoveries.}
\end{figure}

My own view on this is that since fundamental physics has not made
significant discoveries since 1979 (Smolin 2007) then a direct
approach on the fundamental problems may not be the best. If there are
other dimensions, they could as well emerge from some serendipitous
astronomical discovery. When looking at the night sky with the naked
eye, averted vision is often best, and an averted, rather than
directly focused approach to discoveries can often pay off best. We
have to keep observing the Universe in all its detail -- general
astronomical observatories of all wavebands are the way to go.

This `fundamentalist' debate is not settled (e.g. White 2007) and will
continue to play out over the next decade. It is ironic that our
telescopes are remembered mostly for the serendipitous discoveries
they made (Keck Fig,~10, HST) and not for the issues for which the original
science case was made, yet we put most of our efforts into the known
science aspects of the science case for a new telescope. Rather like
the common view of democracy, or peer review, this common approach is
highly flawed but is however the best available.

\bigskip
\noindent{\bf References}

\noindent Bell Burnell J, in Serendipitous Discoveries in Radio
Astronomy, Edited by K Kellermann \& B Sheets, NRAO Greenbank USA,
p160

\noindent Fabian A.C., Signalling over stellar distances with X-rays,
JBIS, 30, 112 (1977)

\noindent Fishman G.J., Harnden F.R., Haymes R.C., Observation of
pulsed hard X-radiation from NP0532 from 1967 data, ApJ, 156, L107 (1969)

\noindent Friedman H,  Astronomer's Universe: Stars, Galaxies and
Cosmos, Ballantine Books, 1991

\noindent Glashow SL, http://physics.bu.edu/static/Glashow/barcelona2002.html 

\noindent Harwit M, Cosmic discovery : the search, scope, and heritage
of astronomy, Basic Books,  1981

\noindent Merton R, Barber E, Travels and Adventures of Serendipity,
Princeton University Press 2006


\noindent Roberts R, Serendipity: Accidental Discoveries in Science,
Wiley 1989
 
\noindent Smolin L. The Trouble with Physics, Penguin 2007

\noindent White S.D.M., Fundamentalist Physics: Why Dark Energy is bad
for Astronomy, Rep. Prog. Phys., 70, 883, 2007

\end{document}